\documentclass[%
 preprint,
 article,
 superscriptaddress,
 showpacs,preprintnumbers,
 amsmath,amssymb,
 aps,
 prc
]{revtex4-1}

\usepackage{graphicx}
\usepackage{dcolumn}
\usepackage{bm}

\usepackage[draft]{pgf}

\begin{document}

\title{Branching ratios in the $\bm{\beta}$ decay of $\bm{^{16}}$N}

\author{O. S. Kirsebom}
\email[Corresponding author: ]{oliver.kirsebom@dal.ca}
\affiliation{Department of Physics and Astronomy, Aarhus University, Denmark}
\affiliation{Institute for Big Data Analytics, Dalhousie University, Canada}

\author{E. R. Christensen}
\email{esbenrohanc@gmail.com}
\affiliation{Department of Physics and Astronomy, Aarhus University, Denmark}

\author{the IS605 Collaboration}

\date{\today}

\begin{abstract}

In this brief note, we present the results of an experiment performed 
at the ISOLDE Decay Station at CERN in which several of the branching ratios in the 
$\beta$ decay of $^{16}$N were determined with high precision and accuracy.

\end{abstract}

\maketitle

The $\gamma$-ray spectrum of $^{16}$N was measured at ISOLDE using an array of 
HPGe detectors, as previously reported in Ref.~\cite{kirsebom2018}. 
The $\gamma$-ray peaks were well resolved, allowing the 
number of counts in each peak to be determined with high precision. 
Yields of individual $\gamma$-ray transitions were determined by dividing 
the number of counts with the energy-dependent detection efficiency of 
the array. Yields were further corrected for summation of $\gamma$ rays 
emitted in cascade transitions, taking into account the known angular 
correlations~\cite{vermeer1982}. Summation was found to be significant 
only for the 8.87-MeV peak, where it accounts for 90\% of the observed yield.

The $\gamma$-ray yields obtained in the present study are given 
in Table~\ref{tb:gamma_intensities}. For unobserved transitions we 
report upper limits at 95\% C.L. The $\beta$-decay branching ratios 
are given in Table~\ref{tb:beta_br}. Finally, the branching 
ratios for the de-excitation of the 8.87-MeV level are given in 
Table~\ref{tb:gamma_br}.
The present results generally have better precision than existing 
literature values~\cite{tunl16}.
Where previous data exist, the agreement with the present data 
is good or fair with one notable exception:
The present value for the branching ratio of the $8.87 \rightarrow 0$ transition 
(Table~\ref{tb:gamma_br}) is a factor of $\sim 60$ smaller than the value reported 
in Ref.~\cite{wilkinson1968}. 
The likely explanation for this discrepancy is that the authors of Ref.~\cite{wilkinson1968} 
failed to account for the summation of $\gamma$ rays emitted in cascade transitions, thereby incorrectly  
attributing the entire yield of the 8.87-MeV peak to the $8.87 \rightarrow 0$ 
transition. Given the information provided in Ref.~\cite{wilkinson1968} it is possible 
to obtain a rough estimate of the summation yield, which in fact is comparable to the 
observed yield.

\begin{table}
\caption{\label{tb:gamma_intensities} $\gamma$-ray yields relative to 
the $6.13\rightarrow 0$ transition. The error bars include the systematic 
uncertainty on the efficiency calibration and statistical uncertainties 
(combined in quadrature).}
\begin{ruledtabular}
\begin{tabular}{ll}
Transition  &  $I_{\gamma}/I_{\gamma_{6.13}}$ (\%)  \\
\hline
$8.87 \rightarrow 0$  &  $0.0021(10)$ \\
$8.87 \rightarrow 6.05$  &  $<0.003$ \\
$8.87 \rightarrow 6.13$  &  $1.52(5)$ \\
$8.87 \rightarrow 6.92$  &  $0.077(3)$ \\
$8.87 \rightarrow 7.12$  &  $0.188(7)$ \\
$7.12 \rightarrow 0$     &  $7.67(8)$ \\
$7.12 \rightarrow 6.05$  &  $<0.006$ \\
$7.12 \rightarrow 6.13$  &  $0.0081(13)$ \\
$6.92 \rightarrow 0$     &  $0.139(3)$ \\
$6.92 \rightarrow 6.05$  &  $<0.004$ \\
$6.92 \rightarrow 6.13$  &  $<0.003$ \\
$6.13 \rightarrow 0$  &  $100$ \\
\end{tabular}
\end{ruledtabular}
\end{table}

\begin{table}
\caption{\label{tb:beta_br} $\beta$-decay branching ratios, including a 0.9\% 
relative uncertainty on the overall normalisation.}
\begin{ruledtabular}
\begin{tabular}{ll}
Level  &  $b_{\beta}$ (\%)  \\
\hline
8.87  &  $1.19(4)$ \\
7.12  &  $5.02(7)$ \\
6.92  &  $0.041(14)$ \\
6.13  &  $66.0(6)$\\
\end{tabular}
\end{ruledtabular}
\end{table}

\begin{table}
\caption{\label{tb:gamma_br} $\gamma$-decay branching ratios of the 8.87-MeV level.}
\begin{ruledtabular}
\begin{tabular}{ll}
Level  &  $\Gamma_{\gamma_{i}} / \Gamma_{\gamma}$ (\%)  \\
\hline
0     &  $0.12(6)$ \\
6.05  &  $<0.2$\\
6.13  &  $85.0(4)$\\
6.92  &  $4.3(2)$ \\
7.12  &  $10.6(5)$ \\
\end{tabular}
\end{ruledtabular}
\end{table}

\bibliography{refs}

\end{document}